\documentclass{article}
\usepackage{url}
\usepackage{graphicx}
\usepackage{amsmath,amsfonts,amssymb}
\usepackage{epsfig,times,latexsym,enumerate,lscape}
\usepackage{amsmath,amssymb,amsfonts}
\usepackage{graphicx}
\usepackage{textcomp}

\begin{document}
\title{\large\bf Secure and Energy-Efficient Key-Agreement Protocol for Multi-Server Architecture
}
\author{Trupil Limbasiya\footnote{BITS, Pilani, Dept. of CS \& IS, Goa Campus, Goa, India, Email: p20170417@goa.bits-pilani.ac.in} \hspace{0.15mm} Sanjay K. Sahay\footnote{BITS, Pilani, Dept. of CS \& IS, Goa Campus, Goa, India, India, Email: ssahay@goa.bits-pilani.ac.in}}

\date{}

\maketitle

\begin{abstract}
	Authentication schemes are practiced globally to verify the legitimacy of users and servers for the exchange of data in different facilities. Generally, the server verifies a user to provide resources for different purposes. But due to the large network system, the authentication process has become complex and therefore, time-to-time different authentication protocols have been proposed for the multi-server architecture. However, most of the protocols are vulnerable to various security attacks and their performance is not efficient. In this paper, we propose a secure and energy-efficient remote user authentication protocol for multi-server systems. The results show that the proposed protocol is comparatively $\sim$44 \% more efficient and needs $\sim$38 \% less communication cost. We also demonstrate that with only two-factor authentication, the proposed protocol is more secure from the earlier related authentication schemes.
\vspace*{0.1cm}
~\\
{\it Authentication, Energy, Multi-server, Security.}
\end{abstract}

\section{Introduction}
In today's emerging world, the Internet has become more and more popular for its various facilities, and it is extensively used in government organizations, smart city applications, education sectors, business, private sectors, etc. Further, there are many applications in which users should get diverse services from different systems remotely, e.g., banking system, healthcare, smart agriculture, smart grid, home automation, etc. Consequently, the network has become highly sophisticated and demanding. Hence, it is not easy to fulfill all users' requirements at the same time, and it leads to the provision of multi-server based system through which applicants can get services any time without any interruption. 

\begin{figure}[h!] 
	\centering
	\includegraphics[width= 90mm]{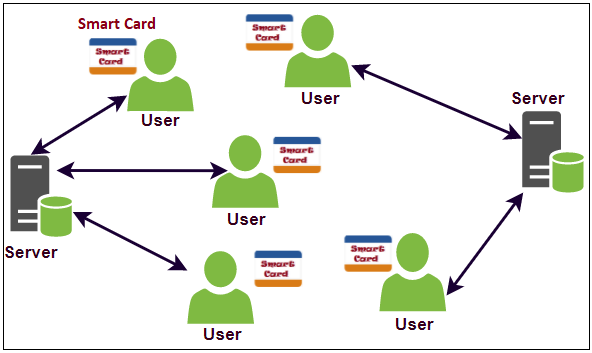}
	\caption{User-server mutual authentication overview in multi-server based structure}
	\label{BASIC}
\end{figure}

In general, server and user should authenticate each other before transmitting/delivering resources to prevent various attacks in a public environment. Therefore, a secure and efficient authentication scheme is required to confirm the legitimacy of the server and a user. In most of the systems, both (user and server) practice the authentication process before starting a communication \cite{Lamport1981}, \cite{Messerges2002}, \cite{Madhusudhan2012}, \cite{Limbasiya2017}. Fig. \ref{BASIC} shows a user-server connection representation for a multi-server environment in which various users are connected to different servers to obtain facility. These servers are typically synchronized with each other, and a user has a smart card to get authenticated by the server. In this scheme, a user cannot connect to multiple servers at the same time, but he/she can establish a connection with different servers alternatively as shown in Fig. \ref{BASIC}.

\subsection{Related Works}
Li et al. \cite{Li2001} suggested an authentication system based on an artificial neural networks to resist a replay attack for the multi-server architecture. Lin et al. \cite{Lin2003} recommended a new authentication protocol to protect a modification attack and a replay attack. Tsaur et al. \cite{Tsaur2005} noticed that an off-line password guessing is feasible in their previous scheme and thus, they designed an improved authentication method to prevent an off-line password guessing attack. However, they (\cite{Li2001}, \cite{Lin2003}, and \cite{Tsaur2005}) used the Diffie–Hellman key exchange concept for the encryption and decryption. Therefore, these schemes are vulnerable to man-in-the-middle, impersonation attacks. Besides, they requires high computational time to perform all necessary operations.

Juang \cite{Juang2004} suggested a mutual authentication and key agreement system for the multiple server framework with low computation and communication cost. Besides, this scheme has various merits, i.e., only one-time registration, no need of a verification table, freely chosen password by a user, mutual authentication, low communication and computation cost. However, they used the symmetric key concept to design an authentication protocol \cite{Juang2004} and thus, there is a key challenge to share the encryption/decryption key between the server and a user. Besides, the encrypted secret key is saved in a smart card. For all these reasons, the scheme \cite{Juang2004} is vulnerable to smart card lost, replay, impersonation, and man-in-the-middle attacks. 

Liao and Wang \cite{Liao2009} proposed an authentication protocol to resist distinct attacks (replay, insider, server spoofing, and stolen verifier). However, Hsiang and Shi \cite{Hsiang2009} found a server spoofing attack is feasible in \cite{Liao2009} as discussed in \cite{Hsiang2009}. To overcome server spoofing and session key attacks, they \cite{Hsiang2009} suggested an enhanced remote user authentication method for the multi-server structure. In 2014, Lee et al. \cite{Lee2014} came up with an extensive chaotic based authentication mechanism to prevent various attacks (plain-text, insider, impersonation, and replay). However, it is vulnerable to denial of service and session key attacks. Besides, both (server and user) need to exchange messages three times to establish a connection for services. Subsequently, Banerjee et al. \cite{Banerjee2015} observed that smart card lost and user impersonation attacks are present in earlier authentication schemes. Then, they proposed a smart card-based anonymous authentication system to prevent security attacks, e.g., user impersonation, smart card lost, forward secrecy, and insider. In 2016, Sun et al. \cite{Sun2016} noticed some loopholes in \cite{Banerjee2015}, i.e., smart card lost, replay, user impersonation, session key, and password guessing. Accordingly, they advised an authentication mechanism using dynamic identity to protect against various attacks (user impersonation, replay, insider, smart card lost, session key, and password guessing). However, we identify that the scheme \cite{Sun2016} is still vulnerable to some attacks, e.g., smart card lost, off-line password guessing, and replay.

Li et al. \cite{Li2016} found security concerns (no single registration, no password update support, and spoofing) in \cite{Lee2014} and proposed a chaotic based key-agreement scheme for enhancing security features. However, Irshad et al. \cite{Irshad2018} found security drawbacks (password guessing, stolen smart card, and user impersonation) in \cite{Li2016}. In addition, they proposed an advanced system to resist identified security concerns in \cite{Irshad2018}. However, Irshad et al.'s scheme \cite{Irshad2018} requires high computational time, storage cost, and communication overhead. Jangirala et al. \cite{Jangirala2017} noticed some security issues in earlier system and advised an extended authentication protocol to enhance security. They also stated that the scheme \cite{Jangirala2017} is resistant to multiple attacks (password guessing, stolen smart card, replay, man-in-the-middle, server spoofing, and forgery), but this scheme is weak in performance. Recently, Ying and Nayak \cite{Ying2019} suggested a remote user authentication mechanism for multi-server architecture using self-certified public key cryptography to improve performance results, but this protocol is susceptible to different attacks, i.e., smart card lost, impersonation, replay, password guessing, session key disclosure, and insider. Moreover, comparatively it requires more computational resources.

\subsection{Contributions}
From the literature survey (\cite{Lee2014} - \cite{Ying2019}), we notice that most of authentication methods are vulnerable to different security attacks and they need more computational resources for the implementation. Thus, we understand that a secure and efficient remote user verification protocol is required for the multi server-based system to provide on-time services and to resist against various security attacks. Therefore, we propose an energy-efficient and more secure remote user authentication scheme, and our contributions are as follows in this paper.  

\begin{itemize}
	\item Design an advanced energy-effective mutual authentication protocol.
	\item Security discussions to check strengths against different attacks, e.g., password guessing, replay, impersonation, insider, session key disclosure, smart card lost, and man-in-the-middle.
	\item Present performance analysis for the proposed method and do the comparison with relevant authentication schemes for different performance measures.
\end{itemize}

The paper is structured as follows. In section 2, we explain the system architecture and the adversary model is described in section 3. In section 4, we propose an advanced authentication protocol using a smart card for multi-server based system. Section 5 discusses performance and security analysis of the suggested system. Then, we do a comparison of the suggested protocol with other related authentication schemes in terms of security and performance. We summarize our conclusions in section 6.

\section{System Architecture}
\begin{figure}[h!] 
	\centering
	\includegraphics[width= 90mm]{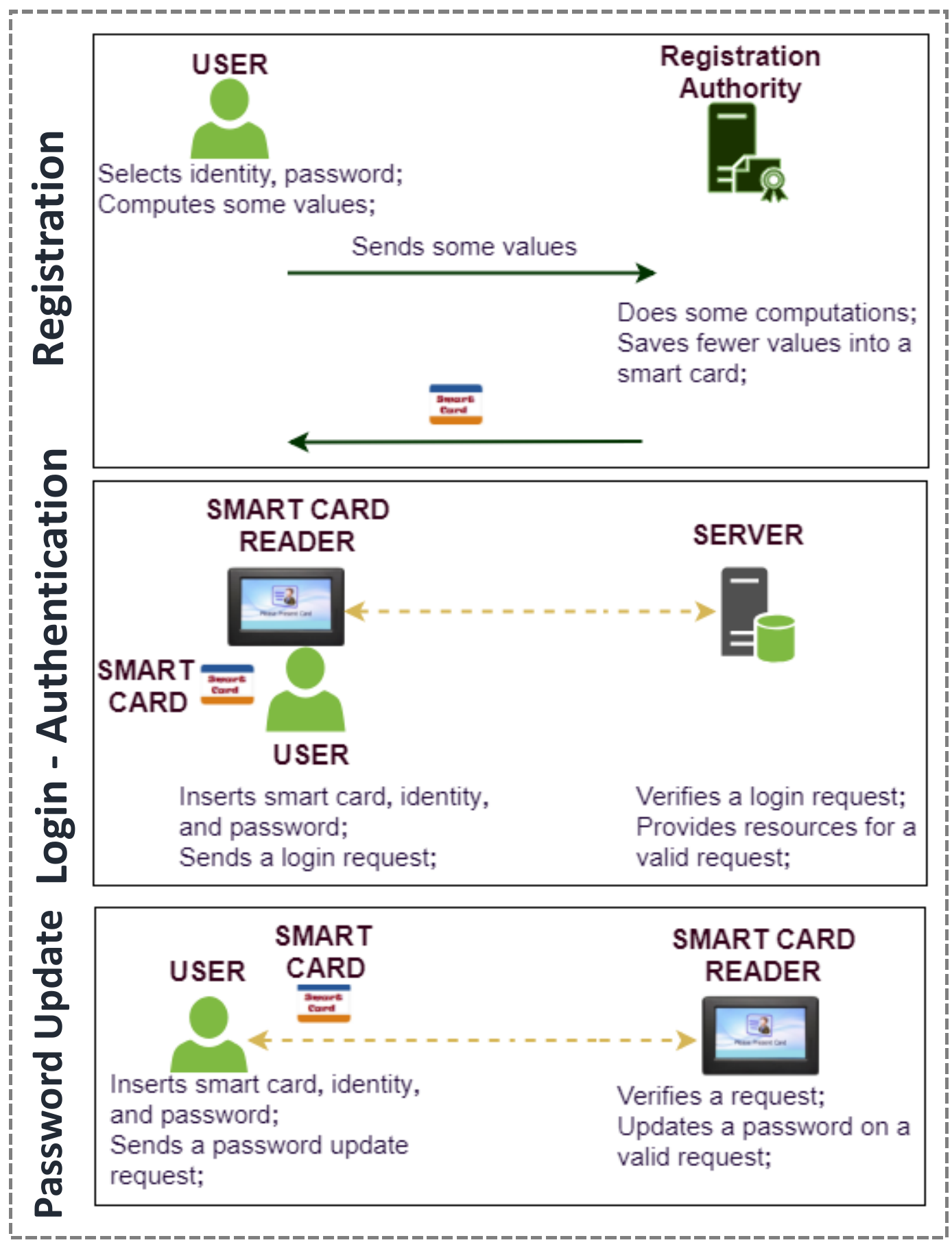}
	\caption{The system architecture}
	\label{SA}
\end{figure}

The registration authority (RA), smart cards, smart card readers, and servers are components in the multi-server system. Users access resources from the server after proving their legitimacy and this data access can be carried out using a smart card through a smart card reader. The RA is a trusted authority in order to register a new user and to provide a legal smart card to that user. Servers are used to provide resources to the legal users and it is highly configured in security, processing power, and storage. A smart card is used to store some important values and these parameters help its owner to get resources/services from the server. In general, a smart card is used to establish a secure connection between a user and the server or to update a user password. The multi-server architecture based system is classified into three phases as (1) registration (2) login and authentication (3) password update. The registration phase is initiated by a new user to become a legitimate person of the system via a secure medium. The login and authentication phase is executed in the interest of a user to access resources from the server over a public channel. The password update phase is performed to change/update a user password via an insecure channel. Fig. \ref{SA} shows an extensive system model overview for different phases by involving different actors.

\section{Adversary Model}
We consider an adversary model according to \cite{Messerges2002}, \cite{Madhusudhan2012}, \cite{Limbasiya2017} for mutual authentication system between a user and the server in a public environment. Accordingly, an attacker has the following capabilities.
\begin{itemize}
	\item An adversary can read/delay/re-transmit packets (transferred over a public channel).
	\item An attacker has the ability to extract parameters from a smart card. And this is feasible after stealing a smart card or getting a lost smart card.
	\item An adversary can modify messages, which are transmitted through an insecure medium.
	\item An attacker can send a forged login request in a polynomial time.
\end{itemize}

\begin{table}[!h]
\centering
\caption{List of different symbols}
\label{tab1}
\begin{tabular}{ll}
\hline
Notations & Explanations\\
\hline
$U_i$ & A user $i$\\
$ID_{i}$ & $U_{i}$'s identity\\
$PW_{i}$ & $U_{i}$'s password\\
$S_{j}$ & A server $j$\\
$SC_{i}$ & $U_{i}$'s smart card\\
$SCR_{k}$ & A smart card reader $k$\\
$N_{i}/N_{j}/b/p_i$ & Random nonce \\
$x/y$ & A server's secret key\\
$y_{i}$ & $U_{i}$'s secret key generated by \textit{RA}\\
$SK_{U_{i}}$ & A session key at $U_{i}$ end\\
$SK_{S_{j}}$ & A session key at $S_{j}$ side\\
$List_{ID_{i}}$ & The list of user identities\\
$List_{F_{i}}$ & The list of computed values $F_{i}$ \\
$\mathcal{A}$ & An adversary/attacker\\
$\oplus$ & Bit-wise XOR operation\\
$\Vert$ & Concatenation operation\\
$h(\cdot)$ & One-way hash function \\
$\Delta T_a$ & A threshold delay fixed at time $a$ \\
$T_b$ & Generated time-stamp at time $b$\\
\hline
\end{tabular}
\end{table}

\section{The Proposed Scheme}
We suggest an energy-efficient remote user authentication protocol for multi-server based system to resist various security attacks. The proposed scheme consists of four phases, (1) server registration, (2) user registration, (3) login and authentication, and (4) user password update as follows.

\subsection{Server Registration}
In the multi-server architecture, different servers should be registered with the registration authority (RA) via an online secure channel. The server registration process is as follows.

\begin{enumerate}
	\item A server ($S_{j}$) chooses an identity ($ID_{S_{j}}$) and password ($PWD_{S_{j}}$), and random nonce ($x_{j}$). Then, $S_{j}$ computes $\alpha_{j}=h(ID_{S_{j}}||PWD_{S_{j}}||x_{j})$ and sends \{$ID_{S_{j}}, \alpha_{j}$\} to the RA.
	
	\item The RA confirms the availability of $ID_{S_{j}}$ and if it is, then the RA does $\beta_{j}=\alpha_{j} \oplus ID_{S_{j}}$., and saves $\beta_{j}$, $\alpha_{j}$, $List_{ID_{i}}$, $List_{F_{i}}$ in the $S_{j}$'s secure storage.
	
\end{enumerate}

\subsection{User Registration}
A new user ($U_{i}$) of the system should enroll with the registration authority (RA) once to become a legal user over a secure channel. $U_{i}$ gets a smart card ($SC_{i}$) after completing the registration process successfully and this $SC_{i}$ helps to get logged into the system for services. This phase is also shown in Fig. 3.

\begin{enumerate}
	\item $U_{i}$ selects $ID_{i}$, $PW_{i}$, $p_{i}$ and calculates $A_{i}=p_{i} \oplus h(ID_{i}||PW_{i})$, $B_{i}=h(PW_{i}||p_{i})$. Then, $U_{i}$ sends \{$ID_{i}, A_{i}, B_{i}$\} to the \textit{RA} over a secure channel.
	
	\item The \textit{RA} generates a random nonce (say $q_{i}$ for $U_{i}$) and enumerates $C_{i} = A_{i} \oplus h(q_{i}||x)$, $D_i=C_{i} \oplus B_{i} \oplus ID_{i}$, $E_i=A_{i} \oplus B_{i} \oplus h(q_{i}||x)$, $F_i=ID_{i} \oplus A_{i} \oplus C_{i}$. After that, it saves $A_{i}, D_{i}, E_{i}$, $List_{ID_{S_{j}}}$ into $SC_{i}$ and $F_{i}$ into the database securely. Next, the \textit{RA} sends $SC_{i}$ to $U_{i}$ via a secure medium.
	
\end{enumerate}

\begin{figure}[!h]
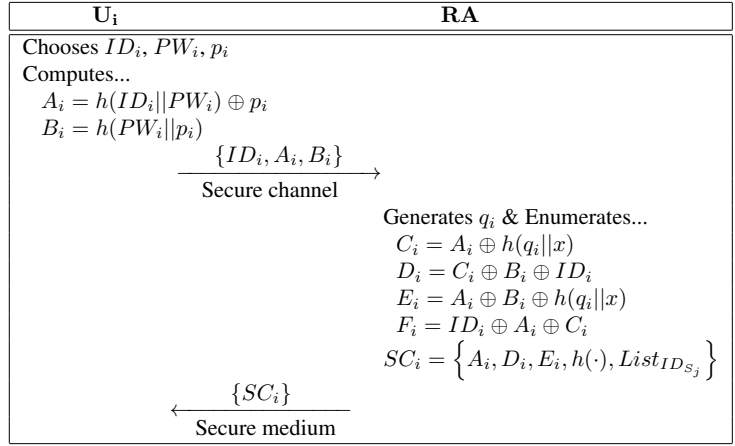

\centering
\label{PRP}
\scalebox{0.85}{
\begin{tabular}{|ll|}
\hline
~$\hspace{1.0cm}\mathbf{U_i}$ & $\hspace{0.5cm}\mathbf{RA}$\\ \hline \hline

Chooses $ID_{i}$, $PW_{i}$, $p_{i}$&\\
Computes...&\\
\hspace{0.2cm} $A_i=h(ID_{i}||PW_{i}) \oplus p_{i}$&\\
\hspace{0.2cm} $B_i=h(PW_{i}||p_{i})$&\\

\multicolumn{1}{|l}{\hspace{2.4cm}$\underrightarrow{\hspace{0.6cm}\{ID_{i}, A_{i}, B_{i}\}\hspace{0.6cm}}$}&\\
\multicolumn{1}{|l}{\hspace{2.8cm}Secure channel}& \\

&\hspace{-0.4cm}Generates $q_{i}$ \& Enumerates...\\
&\hspace{-0.2cm}$C_{i} = A_{i} \oplus h(q_{i}||x)$\\
&\hspace{-0.2cm}$D_i=C_{i} \oplus B_{i} \oplus ID_{i}$\\
&\hspace{-0.2cm}$E_i=A_{i} \oplus B_{i} \oplus h(q_{i}||x)$\\
&\hspace{-0.2cm}$F_i=ID_{i} \oplus A_{i} \oplus C_{i}$\\
&\hspace{-0.4cm}$SC_i=\Big\{A_{i}, D_{i}, E_i, h(\cdot), List_{ID_{S_{j}}}\Big\}$\\

\multicolumn{1}{|l}{\hspace{2.3cm}$\underleftarrow{\hspace{0.95cm}\{SC_i\}\hspace{0.95cm}}$}&\\
\multicolumn{1}{|l}{\hspace{2.7cm}Secure medium}&\\

\hline
\end{tabular}}
\caption{The proposed registration phase}
\end{figure}

\subsection{Login and Authentication}
When $U_{i}$ wants to access service(s) from the server ($S_{j}$), this phase is executed between a smart card reader ($SCR_{k}$) and $S_j$. For this, $U_{i}$/$SCR_k$ performs following steps. Fig. 4 presents the proposed login and authentication phase.

\begin{enumerate}
	\item $U_{i}$ puts $SC_{i}$, $ID_{i}$, and $PW_{i}$ into $SCR_k$. Then, $SCR_{k}$ computes $p'_i=h(ID_{i}||PW_{i}) \oplus A_{i}$, $B'_i=h(PW_{i}||p'_{i})$, $C'_i=B'_{i} \oplus D_{i} \oplus ID_{i}$, $h(q_{i}||x)=C'_i \oplus A_{i}$, $E'_i=A_{i} \oplus B'_{i} \oplus h(q_{i}||x)$. Now, $SCR_{k}$ checks the correctness by comparing $E_{i}$ and $E'_i$. If both are equal, $SCR_{k}$ generates $N_{i}$ and enumerates $AID_{i}=ID_{i} \oplus ID_{S_{j}}$, $G_i=N_{i} \oplus C'_{i} \oplus A_{i}$,  $H_i=G_{i} \oplus h(SID_{j}||N_{i}||T_{1})$, $I_i=N_{i} \oplus p'_{i} \oplus ID_{i}$, $J_i=h(G_{i}||H_{i}||I_{i}||p'_{i}||T_{1})$. Then, it sends \{$ID_{i}, G_i, I_i, J_i, T_1$\} to $S_{j}$ publicly.
	
	\item $S_{j}$ confirms the validity of \{$ID_{i}, F_i, H_i, I_i, T_1$\} by calculating $\Delta T_{1} \leq T_2 - T_1$. If it is valid, then it computes $ID_{i}=\beta_{j} \oplus AID_{i} \oplus \alpha_{j}$, $N'_i=F_i \oplus G_{i} \oplus ID_{i}$, $H'_i=G_{i} \oplus h(SID_{j}||N'_{i}||T_{1})$, $p'_i=N'_i \oplus ID_{i} \oplus I_{i}$, $J'_i=h(G_{i}||H'_{i}||I_{i}||p'_{i}||T_{1})$. Further, $S_{j}$ performs $J'_i \stackrel{?}{=} J_{i}$ for the verification. $S_{j}$ continues to the next step in case of equality. Otherwise, it ends the session immediately.
	
	\item $S_{j}$ generates $N_{j}$ and enumerates $K_{j}=N'_{i} \oplus N_{j} \oplus p'_{i}$, $M_{j}=h(K_{j}||N'_{i}||N_{j}||J'_{i}||T_{2})$. Subsequently, it transfers \{$K_{j}, M_{j}, T_{2}$\} to $U_{i}$.
	
	\item $SCR_{k}$ calculates $\Delta T_{2} \leq T_3 - T_2$ to check its freshness. And it computes $N'_j=N_{i} \oplus K_{j} \oplus p'_{i}$, $M'_{j}=h(K_{j}||N_{i}||N'_{j}||J_{i}||T_{2})$ and verifies $M'_{j}$ by comparing with $M_{j}$. If it holds, then $SCR_{k}$ and $S_{j}$ generates a session key as $h(ID_{i}||SID_{j}||N_{i}||N_{j}||p'_{i}||H_{i}||\Delta T_{1})$. Ultimately, $SCR_{k}$ and $S_{j}$ communicate based on this session key for a limited period. 
\end{enumerate}

\subsection{Password update}
The password change is a facility for the system users to update their $PW_{i}$ for different reason(s) later. For this, $U_{i}$ should imitate following steps.

\begin{enumerate}
	\item $U_{i}$ inserts $SC_{i}$, $ID_{i}$, and $PW_{i}$ into $SCR_{k}$. Then, $SCR_{k}$ enumerates $p'_i=h(ID_{i}||PW_{i}) \oplus A_{i}$, $B'_i=h(PW_{i}||p'_{i})$, $C'_i=B'_{i} \oplus D_{i} \oplus ID_{i}$, $h(q_{i}||x)=C'_i \oplus A_{i}$, $E'_i=A_{i} \oplus B'_{i} \oplus h(q_{i}||x)$. Now, $SCR_{k}$ performs $E_{i} \stackrel{?}{=} E'_i$ to confirm equality. If it holds, $SCR_{k}$ asks for a new password ($PW_{New_{i}}$) and computes $A_{New_{i}}=h(ID_{i}||PW_{New_{i}}) \oplus p'_{i}$, $B_{New_{i}}=h(PW_{New_{i}}||p'_{i})$, $C_{New_{i}}=A_{New_{i}} \oplus h(q_{i}||x)$, $D_{New_{i}}=B_{New_{i}} \oplus ID_{i} \oplus C_{New_{i}}$, $E_{New_{i}}=A_{New_{i}} \oplus h(q_{i}||x) \oplus B_{New_{i}}$. After that, $SCR_{k}$ replaces $A_{i}$, $D_{i}$, and $E_{i}$ by $A_{New_{i}}$, $D_{New_{i}}$, $E_{New_{i}}$ into $SC_{i}$. Finally, $U_{i}$ will have updated $SC_{i}$.
\end{enumerate}

\begin{figure}[!t]
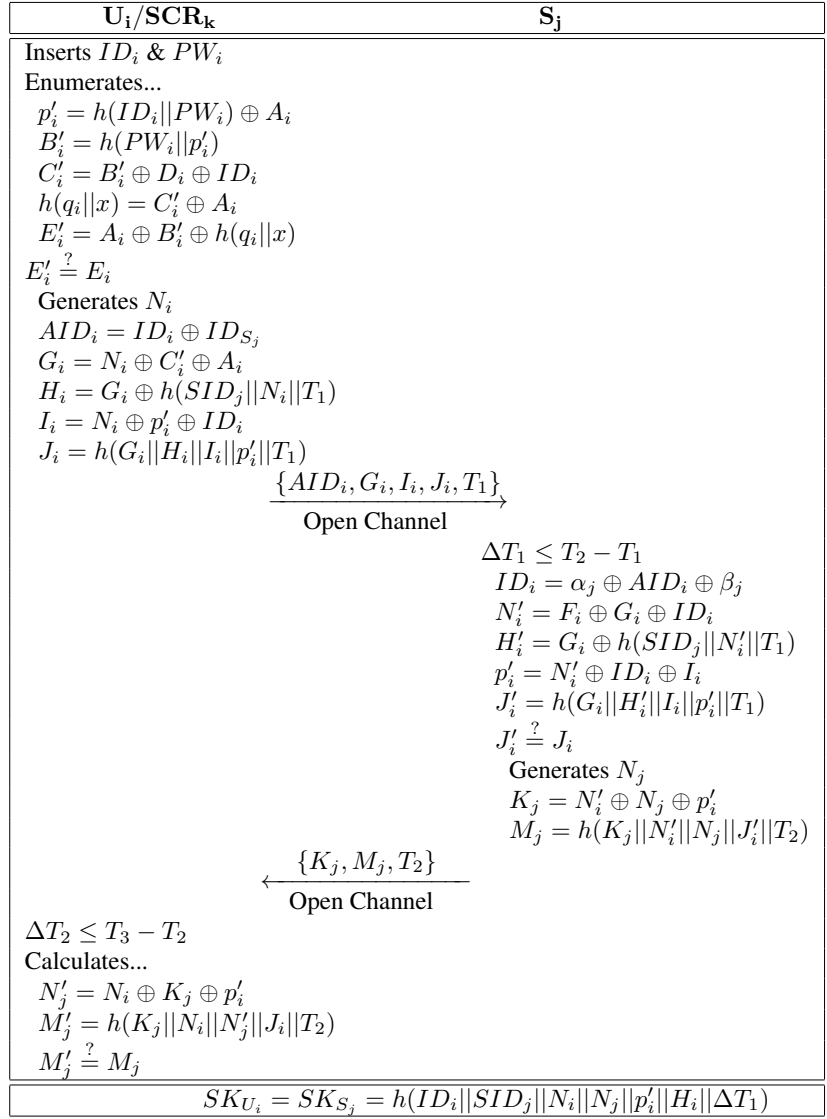

\centering
\label{PLAP}
\scalebox{0.95}{
\begin{tabular}{|ll|}
\hline
~$\hspace{1.0cm}\mathbf{U_{i}/SCR_{k}}$ & $\mathbf{S_{j}}$\\ \hline \hline

Inserts $ID_{i}$ \& $PW_{i}$&\\
Enumerates...&\\
\hspace{0.1cm} $p'_i=h(ID_{i}||PW_{i}) \oplus A_{i}$&\\
\hspace{0.1cm} $B'_i=h(PW_{i}||p'_{i})$&\\
\hspace{0.1cm} $C'_i=B'_{i} \oplus D_{i} \oplus ID_{i}$&\\
\hspace{0.1cm} $h(q_{i}||x)=C'_i \oplus A_{i}$&\\
\hspace{0.1cm} $E'_i=A_{i} \oplus B'_{i} \oplus h(q_{i}||x)$&\\
$E'_i \stackrel{?}{=} E_{i}$&\\

\hspace{0.1cm} Generates $N_{i}$&\\
\hspace{0.1cm} $AID_{i}=ID_{i} \oplus ID_{S_{j}}$&\\
\hspace{0.1cm} $G_i=N_{i} \oplus C'_{i} \oplus A_{i}$&\\
\hspace{0.1cm} $H_i=G_{i} \oplus h(SID_{j}||N_{i}||T_{1})$&\\   
\hspace{0.1cm} $I_i=N_{i} \oplus p'_{i} \oplus ID_{i}$&\\
\hspace{0.1cm} $J_i=h(G_{i}||H_{i}||I_{i}||p'_{i}||T_{1})$&\\

\multicolumn{1}{|l}{\hspace{3.4cm}$\underrightarrow{\hspace{0.1cm}\{AID_{i}, G_i, I_i, J_i, T_1\}\hspace{0.1cm}}$}&\\
\multicolumn{1}{|l}{\hspace{3.9cm}Open Channel}& \\
 
&\hspace{-0.9cm} $\Delta T_{1} \leq T_2 - T_1$\\
&\hspace{-0.7cm} $ID_{i}=\alpha_{j} \oplus AID_{i} \oplus \beta_{j}$\\
&\hspace{-0.7cm} $N'_i=F_i \oplus G_{i} \oplus ID_{i}$\\
&\hspace{-0.7cm} $H'_i=G_{i} \oplus h(SID_{j}||N'_{i}||T_{1})$\\
&\hspace{-0.7cm} $p'_i=N'_i \oplus ID_{i} \oplus I_{i}$\\
&\hspace{-0.7cm} $J'_i=h(G_{i}||H'_{i}||I_{i}||p'_{i}||T_{1})$\\
&\hspace{-0.7cm} $J'_i \stackrel{?}{=} J_{i}$\\

&\hspace{-0.5cm} Generates $N_{j}$\\

&\hspace{-0.5cm} $K_{j}=N'_{i} \oplus N_{j} \oplus p'_{i}$\\
&\hspace{-0.5cm} $M_{j}=h(K_{j}||N'_{i}||N_{j}||J'_{i}||T_{2})$\\

\multicolumn{1}{|l}{\hspace{3.3cm}$\underleftarrow{\hspace{0.5cm}\{K_{j}, M_{j}, T_{2}\}\hspace{0.5cm}}$}&\\
\multicolumn{1}{|l}{\hspace{3.7cm}Open Channel}&\\

$\Delta T_{2} \leq T_3 - T_2$&\\
Calculates...&\\
\hspace{0.1cm} $N'_j=N_{i} \oplus K_{j} \oplus p'_{i}$&\\
\hspace{0.1cm} $M'_{j}=h(K_{j}||N_{i}||N'_{j}||J_{i}||T_{2})$&\\
\hspace{0.1cm} $M'_j \stackrel{?}{=} M_{j}$&\\

\hline
\hline
&\hspace{-4.7cm}$SK_{U_{i}} = SK_{S_{j}} = h(ID_{i}||SID_{j}||N_{i}||N_{j}||p'_{i}||H_{i}||\Delta T_{1})$\\
\hline
\end{tabular}}

\caption{The proposed login and authentication}
\end{figure}

\section{Analysis of the Proposed Scheme}
After proposing an advanced authentication system, we do analysis on this protocol to verify security robustness and performance efficiency. For this confirmation, we have discussed security analysis and performance analysis as below.

\subsection{Security Analysis}
We explain various security attacks and how the proposed system is resistant to different attacks. Then, we compare security robustness of the suggested scheme with other related authentication mechanisms (\cite{Lee2014}, \cite{Banerjee2015}, \cite{Sun2016}, \cite{Li2016}, \cite{Jangirala2017}, \cite{Irshad2018}).

\subsubsection{\textbf{Password Guessing}}
If an adversary ($\mathcal{A}$) can identify the correctness of a guessed password ($PW'_{i}$), then a password guessing attack is possible. $SCR_{k}$ sends \{$ID_{i}, G_i, I_i, J_i, T_1$\} to $S_{j}$ over an open channel. Therefore, $\mathcal{A}$ has access to these parameters. To become a successful in this attack, $\mathcal{A}$ needs to compare $PW'_{i}$ at least with one variable in which $PW_{i}$ has been used and that parameter should be available publicly. $G_{i}$ is computed using $PW_{i}$ indirectly and thus, $\mathcal{A}$ has an opportunity to know correctness of $PW'_{i}$ if he or she can get/derive/compute $G'_{i}$. For this, $\mathcal{A}$ requires $N_{i}$, $C_{i}$, $A_{i}$. However, $N_{i}$ and $p'_{i}$ are generated randomly at $U_{i}$ end, and these random values are only known to $U_{i}$. Therefore, it is difficult to obtain $A_{i}$ and $N_{i}$ exactly. Further, $\mathcal{A}$ does not have essential credentials ($B_{i}$ and $D_{i}$) to compute $C_{i}$. Hence, it is infeasible to derive $G'_{i}$ by having only $PW'_{i}$. Consequently, $\mathcal{A}$ has no opportunity to compare $G'_{i}$. For this reason, $\mathcal{A}$ cannot apply a password guessing attack in the suggested system.

\subsubsection{\textbf{User Impersonation}}
A user impersonation attack is feasible if $\mathcal{A}$ has a favorable plan to create a fake login request, and it should be accepted by $S_{j}$. For this, $\mathcal{A}$ should know or compute $ID_{i}$, $G_i$, $I_i$, $J_i$ in the proposed model. First of all, $T_{1}$ is used in $H_{i}$ and $J_{i}$. Therefore, $\mathcal{A}$ needs to compute these variables ($J_{i}$ and $H_{i}$) to forge a login request. As a result, s/he requires some amount of time to forge a request or generate a fake login request in future. Accordingly, s/he should use fresh time-stamp (say $T'_{1}$). In order to work out for forged parameters ($J'_{i}$ and $H'_{i}$), $\mathcal{A}$ needs $G_{i}$, $N_{i}$, $p'_{i}$. Here, $G_{i}$ is calculated as $N_{i} \oplus A_{i} \oplus C'_{i}$ and hence, $\mathcal{A}$ should know $A_{i}$ and $C'_{i}$ additionally. In the proposed method, $N_{i}$ and $p'_{i}$ are randomly generated numbers. Moreover, $\mathcal{A}$ is not able to enumerate $C'_{i}$ and $A_{i}$ due to unavailability of essential credentials ($B_{i}$, $p'_{i}$, $PW_{i}$, and $D_{i}$). For these reasons, an adversary cannot obtain required credentials anyhow and thus, s/he is restricted to forge $H_{i}$ and $I_{i}$. Additionally, $S_j$ confirms the validity of a login request. As a result, $\mathcal{A}$ fails to make feasible a user impersonation attack in the proposed method.

\subsubsection{\textbf{Replay}}
In the proposed scheme, we have used the concept of a time-stamp to identify transaction time. Here, $SCR_{k}$ sends \{$ID_{i}, J_i, G_i, I_i, T_1$\} to $S_{j}$ over an open medium and $S_{j}$ transfers \{$M_{j}$, $K_{j}$, $T_{2}$\} to $SCR_{k}$ via an insecure channel. Thus, $\mathcal{A}$ can attempt to stop or delay this request/response. However, $S_{j}$ confirms validity of \{$ID_{i}, G_i, I_i, J_i, T_1$\} by executing $\Delta T_{1} \leq T_{2} - T_{1}$. Similarly, $SCR_{k}$ proceeds further after verifying (by calculating $\Delta T_{2}$) the reasonableness of \{$M_{j}$, $K_{j}$, $T_{2}$\}. If $SCR_{k}$ does not get a response message from $S_{j}$ within a reasonable time, then $U_{i}$ understands that $\mathcal{A}$ has tried to interrupt \{$ID_{i}, J_i, I_i, G_i, T_1$\} and after that, $SCR_{k}$ terminates the session directly. It means that if $\mathcal{A}$ tries to perform a replay attack, it will be identified at the receiver side. Additionally, $\mathcal{A}$ cannot change $T_{1}$ in the request or $T_{2}$ in the response message because these time-stamps are used in $H_{i}$, $J_{i}$, $M_{j}$ and these parameters are confirmed at the receiver end. Furthermore, $\mathcal{A}$ does not have essential credentials to calculate $H_{i}$, $J_{i}$, $M_{j}$. After these considerations, an adversary is not able to perform a replay attack in the advised protocol.

\subsubsection{\textbf{Smart Card Lost}}
This attack is applicable if $\mathcal{A}$ can deal with $S_{j}$ mutually and successfully after sending a bogus login request. We assume that a legitimate user ($U_{i}$) can lose his/her $SC_{i}$ or someone can steal $SC_{i}$. Therefore, $\mathcal{A}$ has knowledge of $SC_{i}$ variables ($A_{i}$, $D_{i}$, $E_{i}$) and common channel parameters ($ID_{i}$, $T_2$, $G_i$, $T_{1}$, $I_i$, $J_i$, $K_j$, $M_j$) according to the suggested protocol for this attack. Here, $\mathcal{A}$ should compute a login request in such a way on which $S_{j}$ should be agreed to process further. $U_{i}$ sends \{$ID_{i}, J_i, I_i, G_i, T_1$\} to $S_{j}$ as a login message. These values are calculated as $J_i=h(G_{i}||H_{i}||I_{i}||p'_{i}||T_{1})$, $I_i=N_{i} \oplus p'_{i} \oplus ID_{i}$, $H_i=G_{i} \oplus h(SID_{j}||N_{i}||T_{1})$. Now, $\mathcal{A}$ needs $G_{i}$, $N_{i}$, $p'_{i}$ for enumerating $H'_{i}$ and $J'_{i}$. But $\mathcal{A}$ does not find any proficiency to obtain/calculate these credentials without knowing $B_{i}$, $p'_{i}$, $PW_{i}$, $N_{i}$, and $D_{i}$ (see Section 5.1.2). Additionally, $S_{j}$ checks freshness of a received login request. If $(T_{2}-T_{1})$ is beyond $\Delta T_{1}$, then $S_{j}$ discards that request immediately. Hence, $\mathcal{A}$ cannot proceed to generate a valid login request and this stops to an adversary for further process. In this fashion, the proposed model is protected against a smart card lost attack.

\subsubsection{\textbf{Session key disclosure}}
If $\mathcal{A}$ can generate/compute a valid session key, then there is a possibility of a session key disclosure attack. A session key is calculated as $SK_{U_{i}} / SK_{S_{j}} = h(ID_{i}||SID_{j}||N_{i}||N_{j}||p'_{i}||H_{i}||\Delta T_{1})$ in the suggested scheme. Thus, $\mathcal{A}$ should know $p'_{i}, \Delta T_{1}, N_{i}, H_{i}$, and $N_{j}$ in order to compute it illegally. We consider that $ID_{i}$ and $SID_{j}$ are identity values of $U_{i}$ and $S_{j}$ and thus, these variables are known to $\mathcal{A}$ generally. Next, $\Delta T_{1}$ is the difference between $T_{2}$ and $T_{1}$. Hence, it is also available to an attacker. However, $\mathcal{A}$ does not have $N_{i}$, $N_{j}$, $p'_{i}$, and $H_{i}$. In the proposed method, $p'_{i}$, $N_{i}$, and $N_{j}$ are random numbers and these are only known to $U_{i}$ and $S_{j}$ for a limited time and for this session only. Further, both ($U_{i}$ and $S_{j}$) are agreed on $N_{j}$ and $N_{i}$ for a fixed period. Accordingly, it is hard to know/get these random values. In the proposed method, $H_{i}$ is computed as $h(SID_{j}||N_{i}||T_{1}) \oplus G_{i}$ and therefore, $\mathcal{A}$ cannot calculate $H_{i}$ correctly without having $N_{i}$. For these reasons, $\mathcal{A}$ is unable to proceed for a session key ($SK_{S_{j}}$ / $SK_{U_{i}}$) anyway. After this analysis, a session key disclosure attack is not feasible in the proposed system.

\subsubsection{\textbf{Man-in-the-middle}}
If a person can understand transmitted request/response messages in public environment, then this attack is considerable. In the proposed mechanism, $SCR_{k}$ transmits \{$ID_{i}, J_i, G_i, I_i, T_1$\} to $S_{j}$ and $S_{j}$ responses to $SCR_{k}$ as \{$M_{j}$, $K_{j}$, $T_{2}$\} through an open channel. Therefore, $\mathcal{A}$ can know $ID_{i}$, $T_{1}$, and $T_{2}$ based on these transactions. $ID_{i}$ is a user's identity and it can be identifiable generally. $T_{2}$ and $T_{1}$ are time-stamps and these time-stamps are not profitable to $\mathcal{A}$ effectively because $T_{1}$ and $T_{2}$ are valid for a limited period only. Accordingly, both ($S_{j}$ and $U_{i}$) do not consider $T_{1}$/$T_{2}$ for further process or do not accept any request/response beyond $\Delta T_{1}$/$\Delta T_{2}$. Next, $\mathcal{A}$ needs other vital credentials (e.g., $PW_{i}$, $p'_{i}$, $N_{i}, h(q_{i}||x)$) to understand $G_{i}$, $J_{i}$, and $I_{i}$. Similarly, $\mathcal{A}$ requires $N_{j}$, $N_{i}$, $h(q_{i}||x)$, $G'_{i}$ for $M_{j}$ and $K_{j}$. But $\mathcal{A}$ cannot obtain these private values based on public channel parameters. Consequently, an attacker fails to work out for a man-in-the-middle attack in the suggested method.

\subsubsection{\textbf{Insider}}
If an authorized user can compute a valid login request using his/her own credentials for another legal user, then an insider attack can be applied in the system. We consider two legitimate users ($U_{A}$ and $U_{i}$) in the proposed scheme and $U_{A}$ acts an adversary to impersonate $U_{i}$. $U_{A}$ has his/her $SC_{A}$ and s/he knows $SC_{A}$ values ($A_{A}, D_{A}, E_{A}$). In general, $U_{i}$ sends a login request to $S_{j}$ via a public channel. Therefore, $U_{A}$ has knowledge of $ID_{i}, J_i, G_i, I_i, T_1$. To get access of system resources behalf of $U_{i}$, $U_{A}$ needs to compute a fake login request freshly and this request should be accepted by $S_{j}$ to generate a session key mutually. For this, $U_{A}$ should enumerate $I_i$, $J_i$, and $G_i$ correctly so that $S_{j}$ will be agreed on these values legitimately. $U_{A}$ should know $N_{i}, ID_{i}, p'_{i}$ (for $I_{i}$), $G_{i}, H_{i}, I_{i}, p'_{i}$ (for $J_i$), and $A_{i}, N_{i}, C_{i}$ (for $G_i$). We have already described that $\mathcal{A}$ cannot get $N_{i}, A_{i}, C_{i}, p'_{i}, H_{i}$ (see Section 5.1.2). Similarly, $U_{A}$ is not able to get relevant credentials. Additionally, a time-stamp is used in the suggested method. Next, $q_{i}$ is a random nonce and it is not known to anyone in the proposed system. Further, $q_{i}$ is concatenated with $x$ and then, $S_{j}$ has performed a one-way hash operation. Accordingly, it is difficult to know $q_{i}$ of $U_{i}$. In this fashion, $U_{A}$ fails to calculate a bogus login request. Thus, the suggested scheme can withstand to an insider attack.

\begin{table}[!h]
\centering
\caption{Security features of various authentications protocols}
  \label{tab2}
  \begin{tabular}{lccccccccc}
    \hline
    & \multicolumn{9}{c}{\textit{Security Attributes}}\\\cline{2-10}
		\textit{Schemes} & A1 & A2 & A3 & A4 & A5 & A6 & A7 & A8 & A9\\
		\hline
		Lee et al. \cite{Lee2014} & $\checkmark$ & $X$ & $X$ & $\checkmark$ & $\checkmark$ & $\checkmark$ & $\checkmark$ & Yes & No \\
		
		Banerjee et al. \cite{Banerjee2015} & $X$ & $X$ & $X$ & $\checkmark$ & $X$ & $X$ & $X$ & Yes & No \\
		
		Sun et al. \cite{Sun2016} & $X$ & $X$ & $X$ & $X$ & $\checkmark$ & $\checkmark$ & $X$ & Yes & No \\
		
		Li et al. \cite{Li2016} & $X$ & $\checkmark$ & $X$ & $X$ & $X$ & $\checkmark$ & \checkmark & Yes & No \\
		
		Jangirala et al. \cite{Jangirala2017} & $X$ & $X$ & $X$ & $X$ & $\checkmark$ & $\checkmark$ & $\checkmark$ & Yes & No \\
		
		Irshad et al. \cite{Irshad2018} & $\bigstar$ & $\checkmark$ & $\bigstar$ & $X$ & $X$ & $\checkmark$ & $\checkmark$ & No & Yes\\
		
		Ying and Nayak \cite{Ying2019} & $X$ & $X$ & $X$ & $X$ & $X$ & $\checkmark$ & $X$ &Yes  & No\\
		
		Proposed & $\checkmark$ & $\checkmark$ & $\checkmark$ & $\checkmark$ & $\checkmark$ & $\checkmark$ & $\checkmark$ & Yes & No \\
		\hline
\end{tabular}
\\
\begin{tabular}{ll}
A1\textbf{:} Smart card lost\textbf{;} A2\textbf{:} Impersonation\textbf{;} A3\textbf{:} Password guessing\textbf{;}\\
A4\textbf{:} Replay\textbf{;} A5\textbf{:} Session key disclosure\textbf{;} A6\textbf{:} Man-in-the-middle\textbf{;} \\
A7\textbf{:} Insider\textbf{;} A8\textbf{:} Two-factor authentication\textbf{;} A9\textbf{:} Three-factor authentication\textbf{;} \\
$\checkmark$\textbf{:} Secure\textbf{;} $X$\textbf{:} Vulnerable\textbf{;} $\bigstar$\textbf{:} Insecure without biometric-identity\textbf{;}
\end{tabular}
\end{table}

Table 2 shows a comparison in terms of different security attributes. A smart card lost attack is feasible in \cite{Banerjee2015}, \cite{Sun2016}, \cite{Li2016}, \cite{Jangirala2017}. Authentication schemes (\cite{Lee2014}, \cite{Banerjee2015}, \cite{Sun2016}) are vulnerable to an impersonation attack. An adversary has an opportunity to confirm a guessed password in \cite{Lee2014}, \cite{Banerjee2015}, \cite{Sun2016}, \cite{Li2016}, \cite{Jangirala2017} easily. The schemes (\cite{Sun2016}, \cite{Li2016}, \cite{Jangirala2017}, \cite{Irshad2018}) cannot withstand against a replay attack. A session key disclosure attack can be performed in \cite{Banerjee2015}, \cite{Li2016}, and \cite{Irshad2018}. Banerjee et al.'s scheme \cite{Banerjee2015} is also weak against a man-in-the-middle attack. A legitimate person acts as an adversary to perform an insider attack in \cite{Banerjee2015} and  \cite{Sun2016}. A biometric identity is used in \cite{Irshad2018} to enhance security but it fails to resist attacks (replay and session key disclosure). Additionally, if we consider that a biometric identity can be forged, then two other attacks (smart card lost and a password guessing) are partially possible in \cite{Irshad2018}. In this way, various authentication schemes (\cite{Lee2014}, \cite{Banerjee2015}, \cite{Sun2016}, \cite{Li2016}, \cite{Jangirala2017}, \cite{Irshad2018}) are insecure against various attacks. However, the proposed scheme can withstand against different security attacks as mentioned in Table 2. Further, the suggested protocol can achieve this security level using two-factor authentication only. Therefore, the proposed method is more secure compared to other schemes.

\begin{table*}[h]
\caption{Execution Cost Comparison for different verification schemes}
    \label{tab3}

  \begin{tabular}{p{1.3in}p{1.6in}p{1.6in}}
    \hline
		\textit{Schemes} & Registration & Login and Authentication\\
		\hline
		Lee et al. \cite{Lee2014} & $3T_{h(\cdot)}$ ($\thickapprox$1.74 \textit{ms}) & $11T_{h(\cdot)} + 6T_{CC}$ ($\thickapprox$132.62 \textit{ms})\\
		
		Banerjee et al. \cite{Banerjee2015} & $7T_{h(\cdot)}$ ($\thickapprox$4.06 \textit{ms}) & $17T_{h(\cdot)}$ ($\thickapprox$9.86 \textit{ms}) \\
		
		Sun et al. \cite{Sun2016} & $6T_{h(\cdot)}$ ($\thickapprox$3.48 \textit{ms}) & $18T_{h(\cdot)}$ ($\thickapprox$10.44 \textit{ms}) \\
 
		Li et al. \cite{Li2016} & $3T_{h(\cdot)}$ ($\thickapprox$1.74 \textit{ms}) & $19T_{h(\cdot)} + 6T_{CC}$ ($\thickapprox$137.26 \textit{ms}) \\

		Jangirala et al. \cite{Jangirala2017} & $7T_{h(\cdot)}$ ($\thickapprox$4.06 \textit{ms}) & $23T_{h(\cdot)}$ ($\thickapprox$13.34 \textit{ms}) \\
		
		Irshad et al. \cite{Irshad2018}	& $3T_{h(\cdot)}$ ($\thickapprox$1.74 \textit{ms}) & $29T_{h(\cdot)} + 6T_{CC}$ ($\thickapprox$143.06 \textit{ms}) \\
		
		Ying and Nayak \cite{Ying2019} & $4T_{h(\cdot)} + 2T_{EC}$ ($\thickapprox$77.76 \textit{ms}) & $9T_{h(\cdot)} + 7T_{EC}$ ($\thickapprox$269.26 \textit{ms})\\
		
		Proposed & $3T_{h(\cdot)}$($\thickapprox$1.74 \textit{ms}) & $10T_{h(\cdot)}$ ($\thickapprox$5.80 \textit{ms}) \\

		\hline
\end{tabular}
\end{table*}

\subsection{Performance Analysis}
We explain different performance measure, i.e., execution time, storage cost, and communication overhead. Then, we present outcomes of various remote user authentication mechanisms based on these performance parameters.

\subsubsection{\textbf{Execution Time}}
It is depended on the total number of needed cryptographic operations to carry out the authentication procedure. In this computational cost, most of the verification schemes have used four different cryptographic functions, e.g., one-way hash ($T_{h(\cdot)}$), elliptic curve cryptography ($T_{EC}$), concatenation ($T_{||}$), chebyshev chaotic ($T_{CC}$), and Ex-OR ($T_{\oplus}$). Generally, these operations expect some amount of time in the execution. We consider the running time based on a specific system configuration, i.e., the Ubuntu 12.04.1 32-bit OS, 2 GB RAM with Intel 2.4 GHz CPU \cite{Irshad2018}. The pairing-based cryptography library is inherited for cryptographic operations. After noting down a running time of these functions, we do not include a computing time for $T_{||}$ and $T_{\oplus}$ because they need highly negligible time to accomplish an operation compared to other functions ($T_{EC}$, $T_{CC}$, and $T_{h(\cdot)}$). Therefore, we consider only $T_{h(\cdot)}$, $T_{EC}$, and $T_{CC}$ for the implementation time and these functions expect 0.58 milliseconds (\textit{ms}), 37.72 \textit{ms} and 21.04 \textit{ms} respectively. The comparability between varied authentication schemes (\cite{Lee2014}, \cite{Banerjee2015}, \cite{Sun2016}, \cite{Li2016}, \cite{Jangirala2017}, \cite{Irshad2018}, and the suggested method) is appeared in Table 3. In general, the registration phase is executed once only but the login and authentication process is performed when a legitimate user wants access system resources. Therefore, we mainly focus on the execution time of the login and authentication phase. After looking requirement of different cryptographic functions in Table 3, Banerjee et al.'s scheme \cite{Banerjee2015} takes less execution time (i.e., 9.86 \textit{ms}) compared to other authentication methods (\cite{Lee2014}, \cite{Sun2016}, \cite{Li2016}, \cite{Jangirala2017}, \cite{Irshad2018}, and \cite{Ying2019}). Next, the scheme \cite{Sun2016} requires 10.44 \textit{ms} to complete the login and authentication phase. Although these protocols (\cite{Banerjee2015} and \cite{Sun2016}) are vulnerable to various attacks (see Table 2). However, the suggested method can be implemented in 5.80 \textit{ms} and it is safe against different security attacks (see Section 5.1). Thus, the proposed protocol can be executed rapidly rather than other mentioned authentication systems.

\subsubsection{\textbf{Storage Cost}}
During the registration or the initialization phase, the registration authority stores some credentials into $SC_{i}$ and for this, the system needs to reserve a specified number of bytes. An identity/random nonce variable needs 8 bytes, a chebyshev chaotic function requires 16 bytes, elliptic curve (EC) needs 64 bytes, and 32 bytes (\textit{SHA-2}) are expected as a storage cost. Lee et al.'s scheme \cite{Lee2014} requires 2 (one-way hash), 1 (time-stamp), and 3 (identity/normal) variables. Banerjee et al.'s model \cite{Banerjee2015} and Sun et al.'s system \cite{Sun2016} need 5 (one-way hash) and 1 (identity) variables individually. Li et al.'s mechanism \cite{Li2016} expects 1 (time-stamp), 3 (one-way hash), and 2 (identity/normal)variables. 5 (one-way hash) and 2 (normal) parameters are needed in \cite{Jangirala2017}. Irshad et al.'s protocol \cite{Irshad2018} requires 3 (one-way hash), 3 (identity/normal), and 1 (chebyshev chaotic) variables. The protocol \cite{Ying2019} requires 3 (EC), 2 (one-way hash), and 1 (random nonce).

However, the suggested system needs to save only four (computed using a one-way hash) and one (normal) parameters. Fig. \ref{CSC} shows required storage memory (in bytes) for various authentication models individually. In general, the system saves different credentials into the smart card once. Consequently, this is a one-time process only. Lee et al.'s scheme \cite{Lee2014} needs 92 bytes in storage, which is $\sim$19 \% less compared to the suggested method but the scheme \cite{Lee2014} is weak to two security attacks (password guessing and impersonation) and it expects very high implementation time (see Table 3).

\subsubsection{\textbf{Communication overhead}}
In the login and authentication procedure, both (sender and receiver) transmits different parameters in order to generate a common session key and therefore, they require to spend essential bytes as a communication overhead. An identity variable needs 8 bytes, EC requires 64 bytes, a chebyshev chaotic function expects 16 bytes, a time-stamp requires 4 bytes, and 32 bytes (\textit{SHA-2}) are needed for a one-way hash during the communication. Lee et al.'s method \cite{Lee2014} can be performed using 3 (chebyshev chaotic) and 5 (one-way hash) parameters. The schemes (\cite{Banerjee2015} and \cite{Sun2016}) require 7 (one-way hash) variables separately. Li et al's system \cite{Li2016} needs 1 (identity), 13 (one-way hash), and 4 (chebyshev chaotic) parameters. The scheme \cite{Jangirala2017} can be carried out with 7 (one-way hash) variables. Irshad et al.'s system \cite{Irshad2018} expects 16 (one-way hash), 1 (identity), and 4 (chebyshev chaotic) parameters. Ying and Nayak's scheme \cite{Ying2019} needs 5 (EC), 1 (identity), and 2 (one-way hash).

\begin{figure}[!h]
\centering
\includegraphics[scale=0.67]{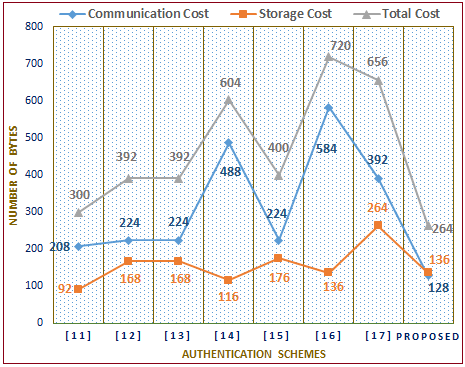}
\caption{Communication and storage cost demand for distinct authentication systems.}
\label{CSC}
\end{figure}

However, the proposed protocol can be implemented using only three (identity), three (one-way hash), and two (time-stamp) variables. The communication cost comparison is shown in Fig. \ref{CSC}. Lee et al.'s scheme \cite{Lee2014} requires 208 bytes for communication but the communication overhead is $\sim$38 \% less in the suggested protocol compared to \cite{Lee2014}. Thus, the proposed method needs less energy to provide services.

\subsubsection{\textbf{Energy Consumption}}
During the authentication phase, the system takes a fixed amount of energy to execute various operations and to send different parameters. This is known as the energy consumption and it is measured in millijoule (mJ). The energy consumption is calculated as $EC_{EXE}=(V*I*t)$ for the execution cost and $EC_{CC}=(V*I*m)/(D_r)$ for communication cost. Where, $V=$ voltage power, $I=$ current, $t=$ the execution time, $m=$ message size, and $D_{r}=$ data rate (6100 Kbps). If the authentication protocol takes low execution time and less communication overhead, then it consumes less energy compared to other authentication schemes. The proposed protocol needs 5.80 \textit{ms} as the execution time and 128 bytes in the communication, which are less compared to other authentication mechanisms (\cite{Lee2014}, \cite{Banerjee2015}, \cite{Sun2016}, \cite{Jangirala2017}, \cite{Irshad2018}, and \cite{Ying2019}). Therefore, the proposed scheme is also energy-efficient compared to other protocols.

\section{Conclusion}
We have proposed a secure and energy-efficient remote user authentication protocol for the multi-server based system. Security analysis of the proposed system is done, and it is shown that our model resists various attacks, i.e., password guessing, impersonation, insider, man-in-the-middle, replay, smart card lost, and session key disclosure even without biometric identity. After analyzing the performance, the results show that the suggested scheme is implemented at least $\sim$44 \% more efficiently compared to relevant schemes. Further, the proposed system comparatively requires $\sim$42 \% less communication overhead and $\sim$19 \% less storage space. Accordingly, the proposed method consumes less energy in the authentication process. To make the multi-server authentication mechanism more attack-proof, we are working  to enhance the security and efficiency of the multi-server authentication system, by analyzing meet-in-the-middle and side-channel attacks, etc.

\end{document}